\documentclass[twocolumn,preprintnumbers,superscriptaddress,nofootinbib,aps,prd,floatfix]{revtex4}
\pdfoutput=1 

\usepackage{epsf,epsfig,subfigure,graphicx,amsmath,amssymb,cancel}
\usepackage{subfigure}
\usepackage{soul}
\usepackage[colorlinks=true,urlcolor=blue,anchorcolor=blue,citecolor=blue,filecolor=blue,linkcolor=blue,menucolor=blue,pagecolor=blue]{hyperref}

\graphicspath{{figures/}}

\newcommand{\be}{\begin{equation}}
\newcommand{\ee}{\end{equation}}
\newcommand{\bea}{\begin{eqnarray}}
\newcommand{\eea}{\end{eqnarray}}
\usepackage{slashed}

\def\beq#1\eeq{\begin{align}#1\end{align}}
\def\beqnn#1\eeq{\begin{align*}#1\end{align*}}

\usepackage[usenames,dvipsnames]{xcolor}
\definecolor{darkgreen}{rgb}{0,0.5,0}
\definecolor{goodorange}{rgb}{0.9,0.4,0}

\begin{document}

\preprint{UCI-HEP-TR-2017-16}

\title{An Update on the LHC Monojet Excess}
\author{Pouya Asadi}
\affiliation{New High Energy Theory Center, Department of Physics and Astronomy,\\Rutgers University, Piscataway, NJ 08854, USA
}
\author{Matthew R. Buckley}
\affiliation{New High Energy Theory Center, Department of Physics and Astronomy,\\Rutgers University, Piscataway, NJ 08854, USA
}
\author{Anthony DiFranzo}
\affiliation{New High Energy Theory Center, Department of Physics and Astronomy,\\Rutgers University, Piscataway, NJ 08854, USA
}
\author{Angelo Monteux}
\affiliation{New High Energy Theory Center, Department of Physics and Astronomy,\\Rutgers University, Piscataway, NJ 08854, USA
}
\affiliation{Department of Physics and Astronomy,\\
University of California, Irvine, CA 92697-4575 USA
}
\author{David Shih}
\affiliation{New High Energy Theory Center, Department of Physics and Astronomy,\\Rutgers University, Piscataway, NJ 08854, USA
}

\begin{abstract}
In previous work, we identified an anomalous number of events  in the LHC jets+MET searches characterized by low jet multiplicity and low-to-moderate transverse energy variables. Here, we update this analysis with results from a new ATLAS search in the monojet channel which also shows a consistent excess. As before, we find that this ``monojet excess" is well-described by the resonant production of a heavy colored state decaying to a quark and a massive invisible particle.  In the combined ATLAS and CMS data, we now find a local (global) preference of 3.3$\sigma$ (2.5$\sigma$) for the new physics model over the Standard Model-only hypothesis. As the signal regions containing the excess are systematics-limited, we consider additional cuts to enhance the signal-to-background ratio. We show that binning finer in $H_T$ and requiring the jets to be more central can increase $S/B$ by a factor of ${\sim} 1.5$.

\end{abstract}

\maketitle


As the LHC reaches a phase of stable running,
it is important to re-examine our search strategies for new physics. Without large increases in energy or luminosity, it becomes less and less likely that new physics will suddenly appear with large statistical significance in a low-background channel. Instead, 
we expect new physics at the LHC to appear only gradually, starting with small deviations from the Standard Model predictions. As the searches for new physics at the LHC grow in sophistication and complexity (especially on the CMS side),  it can become increasingly difficult to separate out statistically-meaningful deviations from random noise. This is exacerbated by the increasing reliance on ``simplified models" to interpret the data. While simplified models are well-suited for limit-setting, they are too few in number (and of too limited variety) to populate more than a small subset of the hundreds of signal regions across all of the LHC searches, so that relying exclusively on simplified models to characterize the data can greatly bias the search for new physics.

In a previous work~\cite{Asadi:2017qon}, we developed a ``rectangular aggregation'' technique which attempted to overcome these biases by combining signal regions in a more model-independent way.  This was based on the simple observation that any signal can populate multiple neighboring bins, and therefore aggregating signal regions within larger kinematic ranges can extract information about underlying excesses without making assumptions about a specific signal model.
As a proof of principle, we applied our aggregation technique to the CMS jets+$\slashed{E}_T$ searches \cite{CMS33} and \cite{CMS36} (hereafter referred to as CMS033 and CMS036, respectively). While originally motivated by supersymmetry, these searches are broadly sensitive to new physics, owing to the fact that they each consist of hundreds of exclusive signal regions, defined by number of jets, number of $b$-tagged jets, and transverse energy variables such as $H_T$, missing transverse momentum $\slashed{E}_T$, and/or $M_{T2}$. 

\begin{figure}[t]
\begin{center}
\includegraphics[width=0.6\columnwidth]{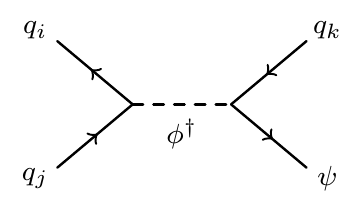}
\caption{The ``mono-$\phi$" simplified model that fits well the monojet excess in the CMS and ATLAS searches.
}
\label{fig:diagram}
\end{center}
\end{figure}

Through our method of rectangular aggregations, we identified a number of interesting ${\sim} 3\sigma$ excesses within these searches. The most interesting one was consistent between both searches. We dubbed this the ``monojet excess'' because it is characterized by low jet multiplicity, no $b$-jets, and low $\slashed{E}_T$ and $H_T$.  
We found that the anomaly's kinematic distributions could not be well-fit by supersymmetry-like pair production of colored particles, or in simplified models for dark matter pair production \cite{Asadi:2017qon}. Instead, a good fit was obtained using a colored scalar $\phi$, resonantly produced through couplings to quarks, and decaying to an invisible massive Dirac fermion $\psi$ and a Standard Model quark (the ``mono-$\phi$'' model), see Fig. \ref{fig:diagram}.

To avoid decays of the $\psi$ back to visible states, its Dirac partner $\psi'$ can be coupled to invisible states $N$ and $\tilde{N}$. The interaction Lagrangian for the minimal model is \cite{Asadi:2017qon}
\begin{equation}
{-\cal L} \supseteq g \phi^* q_i^c \psi + \lambda \phi q_i^c q_j^c + m_\psi \psi\psi' + m_\phi^2 |\phi|^2 + g' \psi' N\tilde{N} + \mbox{h.c.}
\label{eq:lagra}
\end{equation} 
Here, the $q_i$ are the right-handed quarks. The scalar $\phi$ is a color-triplet, and its charge can be $+\tfrac{2}{3}$ or $-\tfrac{1}{3}$. For a given $\phi$ mass, the resonance production cross section is set by $\lambda$, while the branching ratios of $\phi$ to $q\psi$ versus $qq$ are set by both $\lambda$ and $g$. The $\phi$ resembles a squark in $R$-parity violating supersymmetry, though in order to avoid baryon-number-violating decays the $\psi$ cannot be identified with a Majorana neutralino \cite{BARBIERI1986679}.

\begin{table*}[t]
\begin{tabular}{c|cccc|cccc}
Search & $N_j$ & $N_b$ & $H_T$ & $M_{T2},\slashed{E}_T$ & Obs. & Bg. (pre-fit) & Bg. (post-fit) & Best-fit Signal \\ \hline
CMS036 & $1-3$ & 0 & $250-450$ & $200-300$ & 145144  & 137256 $\pm$ 8159  & 140391 $\pm$ 1524 & 4753
\\
CMS033 & $2-4$ & 0 & $300-500$ & $300-500$  & 58138 & 54550 $\pm$ 2246 & 55976 $\pm$ 780 & 2162
\\
ATLAS060& $\geq1$ & - & $\geq250$ & $350-700$ & 74686 & \multicolumn{2}{c}{72645 $\pm$ 1140} & 2041
\\\hline
\end{tabular}
\caption{Regions of kinematic space containing the excess, and relative observed and background event counts, as well as number of signal events after the fit.
For the background, we quote both the pre- and post-fit values for the expected counts and the relative errors.
\label{tab:data}}
\end{table*}

We also found further hints for the same anomaly in the ATLAS 2-6 jets+$\slashed{E}_T$ search  \cite{ATLAS22} (ATLAS022) which, owing to high $m_{\rm eff}$ and $\slashed E_T$ thresholds, was not as sensitive to this signal as it could have been. The anomaly is in some tension with null results from the CMS dark matter+jets exotica search~\cite{CMS48} (CMS048), but a production cross section on the order of 0.5~pb can evade the 95\%CL limits from that search, while still maintaining a local $3\sigma$ preference for signal over background in CMS036 and ATLAS022.\footnote{As the data sets between the searches from a single collaboration are overlapping, we cannot  statistically combine multiple CMS or ATLAS analyses, and must confine ourselves to a single CMS and a single ATLAS result.}

This letter serves as an update to the original analysis, containing three new points concerning the monojet excess:
\begin{enumerate}
\item We include the newly released ATLAS monojet search~\cite{ATLAS60} (ATLAS060). This is a search for a dark matter mediator in events with missing energy and at least one high-$p_T$ jet ($p_T^{j_1}>250$~GeV), with ten exclusive bins in the $\slashed E_T$ variable, starting at $\slashed E_T=250$~GeV. This search has a better sensitivity to the mono-$\phi$ model than ATLAS022, where the most sensitive signal region had much higher thresholds, requiring two jets and $m_{\rm eff}\equiv \slashed E_T+\sum_j p_T^j>1200$~GeV. In the ATLAS060 data, we find a $2-2.5\sigma$ preference for this model in the same region of parameter space preferred by the CMS searches. The previous ATLAS022 analysis had only a $1-1.5\sigma$ preference.  

\item We perform a joint statistical analysis of the CMS and ATLAS data, including the look-elsewhere-effect. Combining the ATLAS060 and CMS036 searches, 
the local significance for this model reaches $3.5\sigma$ in a region of parameter space in the $m_\phi=1200-1800$~GeV range and mass splitting $m_\phi-m_\psi=300-400$~GeV, which is lowered to $3.3\sigma$ when requiring the signal cross section to be allowed at 95\%CL by CMS048.
Using 10,000 pseudoexperiments, we estimate that this corresponds to a $2.5\sigma$ global significance.

\item We suggest additional cuts to enhance the experimental sensitivity to this signal. As we will describe, the experimental errors for the signal regions containing the observed excess are systematics-dominated. Therefore, additional data may not appreciably increase the overall significance of the anomaly, even if it is due to new physics. With the production mode in Fig.~\ref{fig:diagram}, the signal $H_T$ distribution is peaked at the mass difference between $\phi$ and $\psi$, while the background is smoothly falling. In addition, the signal jet tends to be more centrally produced than the background. Therefore, we find the most effective way to increase sensitivity is to define finer $H_T$ bins and require a tighter cut on the leading jet pseudo-rapidity, in particular $|\eta|\lesssim 0.5$. We find that $S/B$ can be increased by a factor of ${\sim}1.5$ compared to the current CMS036 analysis, to an overall level of $S/B\sim 8\%$. 

\end{enumerate}

In Table~\ref{tab:data}, we show the range of kinematic parameters within the various ATLAS and CMS searches containing the anomalous events which we have identified as the monojet excess. For the CMS033 and CMS036 searches, the anomaly is spread out over a number of signal regions.  The uncertainties on the expected number of events in these signal regions are highly correlated, and we make use of simplified covariance matrices provided by CMS \cite{CMS36} (no correlations were provided for ATLAS060). In Table~\ref{tab:data}, we report both ``pre-fit'' and ``post-fit'' background predictions. (We are following standard CMS terminology, see e.g.~\cite{CMS-PAS-HIG-15-002}) The pre-fit backgrounds are the simple aggregation of the background counts and the sum of covariance matrix for the bins populated by signal (as detailed in~\cite{Asadi:2017qon}). The post-fit values refer to a combined fit of all the signal regions in the presence  of a new-physics signal which only populates a specific subset of all the bins. Due to the high degree of correlation between bins populated by the signal and those bins where no signal events fall, post-fit errors are reduced: effectively, the bins not populated by signal act as additional control regions and lower the uncertainty in the bins of interest.\footnote{We thank Claudio Campagnari for emphasizing this point in our procedure.}  In the signal regions of interests, with pre-fit errors of order $4-5\%$, this procedure results in post-fit uncertainties at the $1\%$ level. As background correlations were not released with the ATLAS060 search, there is no difference between pre-fit and post-fit for that search. 

\begin{figure}[t]
\begin{center}
\includegraphics[width=1\columnwidth]{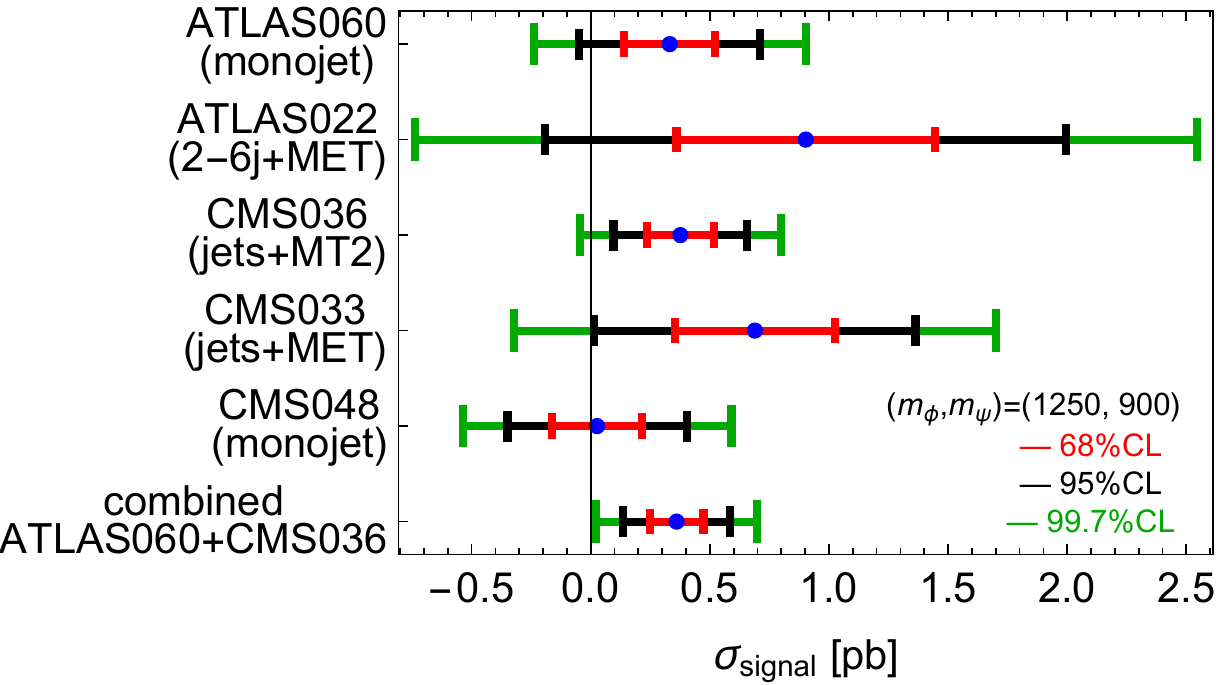}
\caption{Values of the signal cross sections favored at 1,2 and 3$\sigma$ by each individual search considered, and by the combination of ATLAS060 and CMS036. }
\label{fig:signalstrength}
\end{center}
\end{figure}

The signal regions of Table~\ref{tab:data} identify the ``core'' of the identified excess, and are independent of any particular new physics model. However, a full fit -- including all signal regions of each search -- requires both a model and a recasting of the experimental search sensitivity for that model. Scanning over the $(m_\phi,m_\psi)$ mass plane, we generated mock-LHC data for the mono-$\phi$ model using \textsc{MadGraph5} \cite{Alwall:2014hca}, \textsc{Pythia8} \cite{Sjostrand:2014zea} for showering and hadronization, and a tuned implementation of \textsc{Delphes3} \cite{deFavereau:2013fsa} for detector simulation. Events were generated without jet matching, though comparison with matched samples demonstrated that the effect was minimal. Full details of our recasting procedure and cross-checks can be found in~\cite{Asadi:2017qon}. For each ATLAS or CMS search~\cite{CMS33,CMS36,ATLAS22,CMS48}, we calculate the statistical preference for the signal+background hypothesis over background-only using the profile likelihood method \cite{SimplifiedLL,Cranmer1007.1727}, treating the cross section times branching ratio at each mass point as a free parameter in the fit. The results are indicated in Fig.~\ref{fig:signalstrength}, where we show the best-fit confidence intervals for $\sigma\times {\rm BR}$ of a reference mass point $(m_\phi,m_\psi) = (1250,900)$~GeV, for each of the ATLAS and CMS searches of interest. As can be seen, the anomaly seen in ATLAS060 is broadly consistent with that previously identified in the CMS033, CMS036, and ATLAS022 data, and at higher significance than the previous ATLAS search. While the CMS monojets search CMS048 did not see any evidence for new physics, its confidence intervals are entirely consistent with the size of the excess seen by the other searches.

\begin{figure}[t]
\begin{center}
\includegraphics[width=0.9\columnwidth]{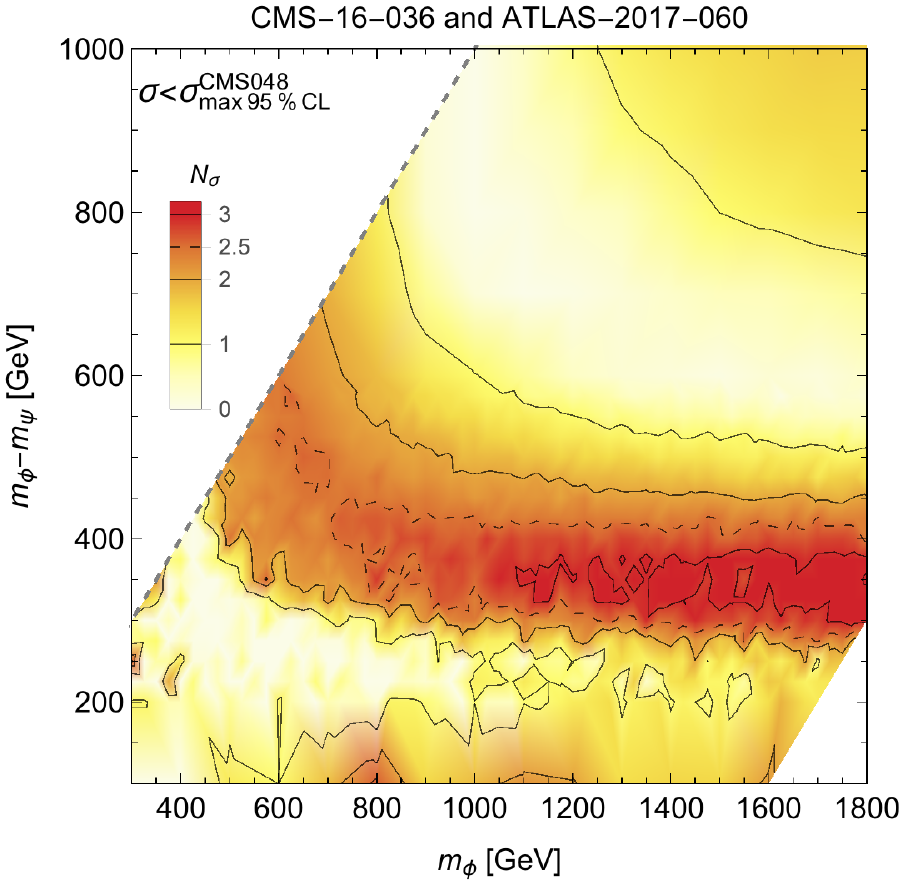}
\caption{Best-fit significance for the model in the $m_\phi/m_\phi-m_\psi$ mass plane, obtained combining the CMS jets+$M_{T2}$ search \cite{CMS36} and the ATLAS monojet search \cite{ATLAS60}. The corresponding best-fit cross section is ${\cal O}(0.35~{\rm pb})$ in the region with highest significance.}
\label{fig:significance}
\end{center}
\end{figure}

Although we cannot combine all of these searches to produce an overall best fit cross section, we can pick one from CMS and one from ATLAS for a joint fit. Choosing CMS036 and ATLAS060 as being the two that are most sensitive to our signal, the resulting significance plot   is shown in Fig.~\ref{fig:significance}. To take into account the non-observation of signal from CMS048, we require that the best-fit cross section be less than the 95\%CL upper limit from that search.\footnote{As discussed in~\cite{Asadi:2017qon}, this model also gives a correlated signature in the dijet resonance channel, however the exact signal strength depends on additional couplings not determined by this fit. Here we assume that the couplings are always chosen such that the $\sigma\times{\rm BR}$ into dijet resonances is consistent with current bounds.} Even after this, the combined fit finds a local preference for signal at the $3.3\sigma$ level  for $m_\phi \sim 1200-1800$~GeV and $m_\phi - m_\psi \sim 300-400$~GeV, with $\sigma\times {\rm BR}\sim 0.3$~pb. This represents an increased significance from the $3\sigma$ result reported in~\cite{Asadi:2017qon}.

\begin{figure}[t]
\begin{center}
\includegraphics[width=1.03\columnwidth]{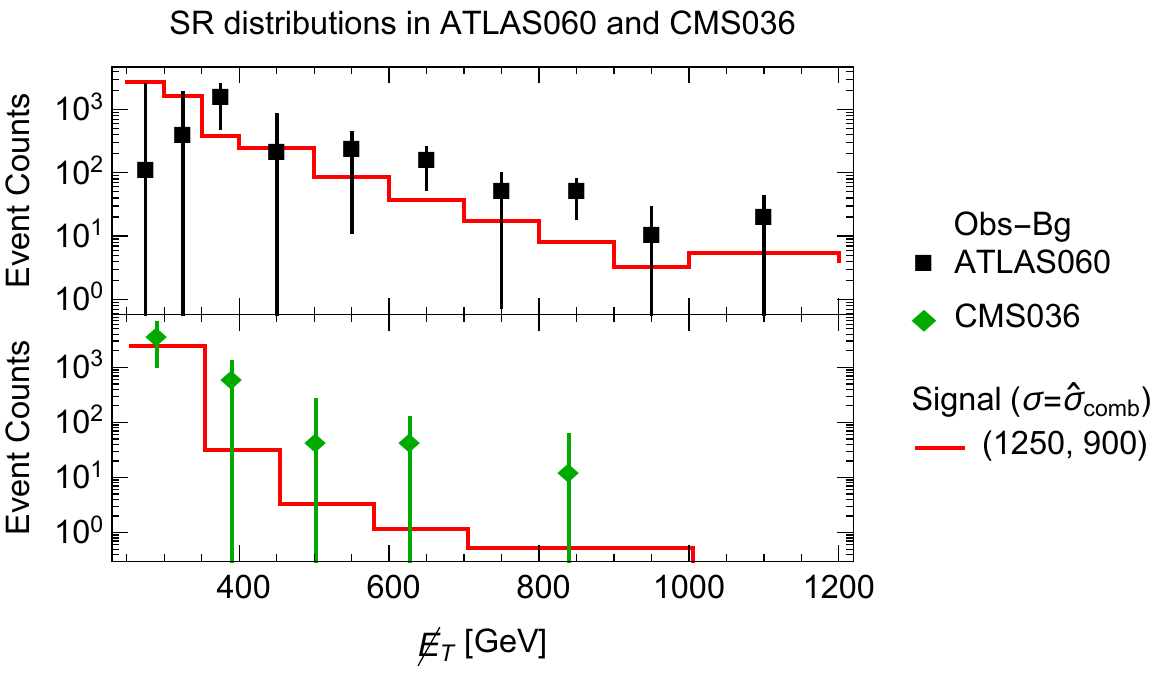}
\caption{Difference between observed and background counts with relative error bars for ATLAS060 (black) and the CMS036 $N_j=1$ bins (green), to be compared with the $\slashed{E}_T$ distribution of the signal for $(m_\phi,m_\psi) = (1250,900)$, respectively in solid and dashed red, given the production cross section set by the joint fit to ATLAS060 and CMS036.}
\label{fig:atlas060_SRdist}
\end{center}
\end{figure}

To additionally illustrate the compatibility between the excesses in CMS036 and ATLAS060, 
in Fig.~\ref{fig:atlas060_SRdist} we show the residuals (observed minus expected) from ATLAS060 and from the $N_j=1$ bins of CMS036 as a function of $\slashed{E}_T$, along with the distribution predicted by the mono-$\phi$ model for the specific mass point $(m_\phi,m_\psi) = (1250,900)$~GeV, using the best-fit cross sections. The kinematics are similar to other good-fit mass points with $m_\phi - m_\psi \approx 300$~GeV.

We further update the analysis of~\cite{Asadi:2017qon} to include the global significance of the combined fit to CMS036 and ATLAS060. We generate 10,000 pseudoexperiments of ATLAS060 and CMS036 data, drawing from the background-only distributions (as in the combined fit, we neglect possible correlations between systematics of ATLAS and CMS). For each pseudoexperiment, we perform a combined fit and count the number of pseudoexperiments for which the significance for the mono-$\phi$ model is greater than observed in the data ($3.3\sigma$), scanning over the mass plane.
The fraction of pseudoexperiments where the background mimics the signal at the $3.3\sigma$ level or more is $0.0128$.
We therefore conclude that our $3.3\sigma$ local excess in the combined dataset corresponds to a global $2.5\sigma$ anomaly in the context of the mono-$\phi$ model.

A $2.5\sigma$ global excess is potentially interesting, but certainly not definitive proof of physics beyond the Standard Model. Since the quoted significance is dominated by systematic errors (see the quoted pre- and post-fit errors in Table~\ref{tab:data}, which are significantly larger than $\sqrt{N}$), the situation might not necessarily improve with more data. Instead, one must either reduce the systematic errors, or identify further cuts that can enhance the new physics scenario over the Standard Model background. Since the former is something only the experimentalists can do, here we focus on the latter.

For specificity, we consider the search with the greatest signal significance (CMS036), though a similar analysis could be performed with ATLAS060. We simulate the primary backgrounds in the relevant signal regions: $(Z\to \nu\nu)$+jets and $(W\to \ell \nu)$+jets, where the lepton is missing.   The events are generated in the same \textsc{MadGraph5}, \textsc{Pythia8}, and \textsc{Delphes3} chain used for our signal, matched up to four jets. We normalize each background sample by reweighting them in each exclusive signal region to match the expected pre-fit background rates reported in CMS036.

\begin{figure*}[tbhp]
\begin{center}
\includegraphics[width=0.66\columnwidth]{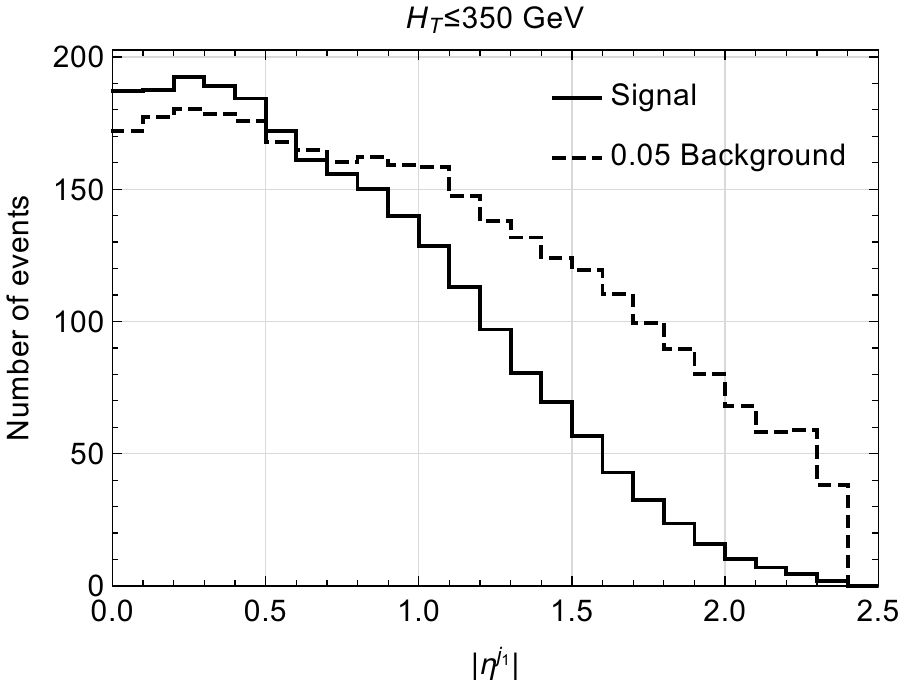}
\includegraphics[width=0.66\columnwidth]{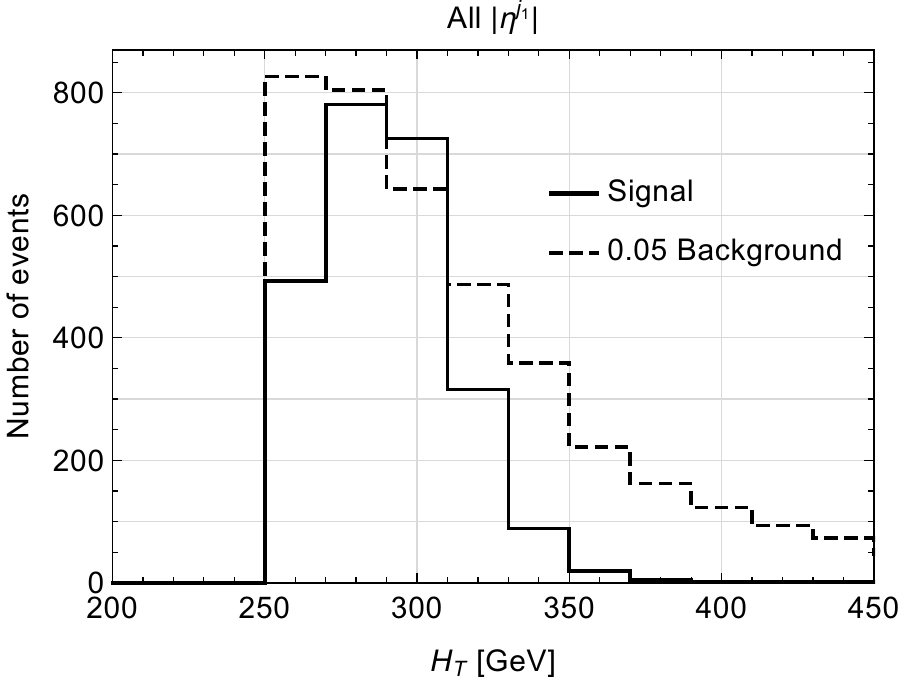}
\includegraphics[width=0.66\columnwidth]{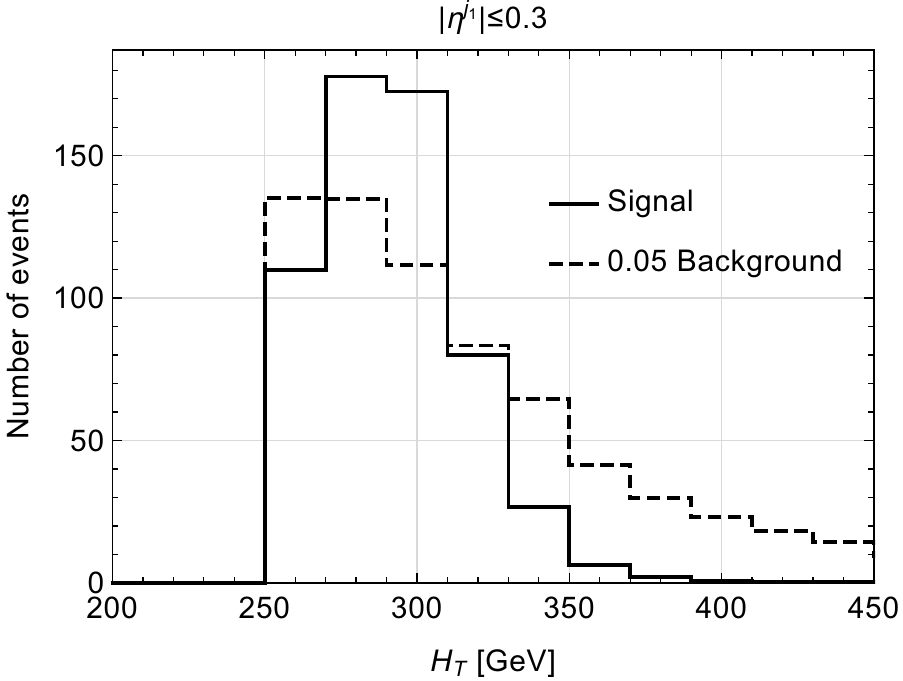}
\caption{Distributions of signal (solid) and  our estimate of the pre-fit systematic error ($5\%$ of the background events) in the $N_j = 1$, $N_b = 0$ SRs of CMS036. Left: Distribution with respect to jet $|\eta|$. Center: Distribution with respect to $H_T$ without a cut on $|\eta|$. Right: Distribution with respect to $H_T$ requiring jet $|\eta| < 0.3$.}
\label{fig:kinplots}
\end{center}
\end{figure*}

Since the signal contains one parton from the hard process, we  focus on the monojet bins  of CMS036, which is defined by the criteria $N_j=1$, $N_b=0$, $H_T\ge 250~{\rm GeV}$.  With only one jet in each event, the only kinematic variables to cut on are the jet $p_T$ and pseudo-rapidity.  As the signal is produced from the decay of a resonantly produced massive scalar, while the background is produced from $t$-channel scale-invariant QCD, we expect the signal to be peaked in $p_T$ and more central than the background.
This is confirmed by the $\eta$ and $H_T$ histograms shown in  Fig.~\ref{fig:kinplots}. There we show the background multiplied by a conservative estimate of the reported pre-fit systematic error (${\sim}5$\%, as can be seen in Table~\ref{tab:data}).

In Fig.~\ref{fig:kinplots}, we further show the $H_T$ distributions before (center panel) and after (right panel) a tighter $\eta$ cut on the jet. It can be seen that by requiring the jet to be more  central, the peak of the signal distribution goes up by a factor of ${\sim} 1.5$ as compared to all events. Moreover, signal and background peak at different values of $H_T$, and so this distribution is possibly robust against systematic errors near threshold. Note that this difference can only be seen if the events are binned in sufficiently small ranges of $H_T$. If, after the inclusion of this additional $\eta$ cut, the ${\sim}5\%$ pre-fit systematic errors can still be reduced to the previously-achieved ${\sim}1\%$ level post-fit, then this factor of $1.5$ boost in signal over the background could potentially increase the local statistical significance from the single CMS036 search up to $\sim 3\sigma \times 1.5 \sim 4.5\sigma$. Similar improvements in signal sensitivity could presumably be performed with the ATLAS data.

In summary, we have updated the analysis described in~\cite{Asadi:2017qon} with the latest monojet search from ATLAS, which appears to have an excess in the same place and is consistent with the excess found in~\cite{Asadi:2017qon}. As in~\cite{Asadi:2017qon},  we find that a resonantly-produced  color-triplet scalar decaying to jet plus missing energy fits all the data well. Performing a joint fit, the local significance of the monojet excess grows to 3.3$\sigma$ local (2.5$\sigma$ global). Finally, we show that by binning more finely in $H_T$ and putting a simple cut on the centrality of the jet, we can enhance signal over background by a factor of at least ${\sim}1.5$. This could greatly increase the significance of the anomaly, as well as providing more confidence that it is not due to systematic errors.

\section*{Acknowledgements}
We thank Claudio Campagnari, Elliot Lipeles and Keisuke Yoshihara for helpful discussions. The work of P.A. and D.S. is supported by DOE grant DE-SC0010008. 
M.R.B.~is supported by DOE grant DE-SC0017811. The work of A.M. was supported in part by NSF Grant No.~PHY-1620638 and in part by Simons Investigator Award \#376204.

\bibliographystyle{utphys}
\bibliography{update}

\end{document}